\documentclass[journal]{IEEEtran}

\usepackage{cite}
\usepackage{amsmath,amssymb,amsfonts}
\usepackage{graphicx}
\usepackage{textcomp}
\usepackage{stfloats}
\usepackage{array}
\usepackage{booktabs}
\usepackage{threeparttable}
\usepackage{multirow}
\usepackage{indentfirst}
\usepackage{bm}
\usepackage[table]{xcolor}
\usepackage[hidelinks]{hyperref}
\usepackage[caption=false,font=normalsize]{subfig}
\usepackage{pifont}
\newcommand{\cmark}{\ding{51}} 
\newcommand{\xmark}{\ding{55}} 

\begin{document}

\title{MiCLA: Mixture-to-Constituent Latent Alignment for Polyphonic Audio Representation Learning}

\author{Kohei~Yamamoto$^{1,2}$, Kosuke~Okusa$^{2}$
\thanks{$^{1}$Research \& Development Center, Technology Division, Oki Electric Industry Co., Ltd., Saitama, Japan. $^{2}$Department of Data Science for Business Innovation, Chuo University, Tokyo, Japan. Kohei~Yamamoto (e-mail: yamamoto833@oki.com) is a Ph.D. student at Chuo University and also with Oki Electric Industry Co., Ltd. Kosuke~Okusa (e-mail: okusa@kc.chuo-u.ac.jp) is an Associate Professor at Chuo University, Tokyo, Japan.}%
}

\maketitle

\begin{abstract}
Real-world audio typically contains concurrent sound events, yet standard self-supervised methods treat each clip as an isolated training unit without explicitly modeling mixture--constituent relationships.
We propose Mixture-to-Constituent Latent Alignment (MiCLA), a two-stage self-supervised pre-training framework that uses synthetic log-mel mixing to align mixture representations with their constituent clips.
Stage~1 structures single-clip representations via masked self-distillation and contrastive regularization.
Stage~2 introduces bidirectional alignment: for a synthetic mixture, MiCLA suppresses the component aligned with one constituent and aligns the residual with the other, using projective residualization on patch tokens and sample-adaptive soft residualization on the global class token.
In downstream evaluations, MiCLA achieves 43.2\% mean average precision (mAP) on AS-20K and 50.3\% mAP on AS-2M under fine-tuning, and achieves the highest mAP among evaluated models on both AudioSet settings under frozen Convex Gated Probing (CGP).
Furthermore, a Joint Semantic Retrieval diagnostic shows that MiCLA improves simultaneous retrieval of both constituent classes over target-averaging baselines.
These results indicate that directional mixture-to-constituent alignment provides an effective pre-training signal for polyphonic audio representation learning.
\end{abstract}


\section{INTRODUCTION}

\IEEEPARstart{R}{eal-world} audio recordings rarely contain only a single isolated sound event.
Even within a short recording, multiple acoustic events, such as speech, machine sounds, music, and background noise, routinely overlap.
Representations learned from such polyphonic audio should ideally capture global acoustic context while preserving constituent-related semantic information in a form accessible to downstream models.
This property is particularly critical for multi-label classification, where multiple co-occurring sound events must be predicted simultaneously.

Recent advances in self-supervised audio representation learning have demonstrated strong transfer performance by leveraging large-scale unlabeled audio data.
Masked audio modeling~\cite{ssast-22,mae-ast-22,audiomae-22,maskspec-23,beats-23}, teacher--student self-distillation~\cite{data2vec-22,data2vec2-23,atst-24,eat-24}, and contrastive learning~\cite{mast-23,slicer-23,audiomosaic-26} learn representations from masked or augmented views.
However, most of these approaches treat each audio clip as a single training unit.
Consequently, these methods do not explicitly constrain how a mixture representation should relate to the representations of the clips used to construct it.
SSLAM~\cite{sslam-25}, a closely related method, supervises masked prediction from a mixed input by averaging the corresponding teacher patch targets from its two pre-mixture inputs.
Such target averaging does not explicitly encourage the model to retain information about both constituent clips in a downstream-accessible manner.

To address this limitation, we propose Mixture-to-Constituent Latent Alignment (MiCLA), a self-supervised pre-training framework for polyphonic audio representation learning.
Rather than averaging the two targets, MiCLA uses the known correspondence between a synthetic mixture and the two pre-mixture clips used to construct it (hereafter referred to as constituent clips) to define two directional residual-alignment tasks.
Given two constituent clips, MiCLA convexly interpolates their log-mel spectrograms to construct a synthetic mixture input.
In one direction, MiCLA suppresses the component of the student mixture representation aligned with the student representation of one constituent and aligns the resulting residual with the EMA-teacher representation of the other.
It then reverses the roles of the two constituents.
This bidirectional formulation is designed to retain information associated with both constituent clips rather than collapsing their supervision into a single averaged target.
Importantly, constituent clips may themselves contain multiple sound events, and MiCLA requires neither class annotations nor isolated source tracks.

MiCLA adopts a two-stage pre-training strategy~\cite{sslam-25} based on a Vision Transformer (ViT)~\cite{vit-21} architecture.
In Stage~1, the ViT model is trained exclusively on constituent clips using teacher--student masked self-distillation~\cite{eat-24} and instance-level contrastive regularization (ICR)~\cite{aasp-26} to establish the constituent-side latent geometry.
Stage~2 then continues pre-training on constituent clips and synthetic mixtures, retaining the Stage~1 objectives to preserve the established constituent-side geometry while adding bidirectional mixture-to-constituent alignment.
Furthermore, MiCLA accounts for the differing representational roles of class (CLS) and patch tokens during bidirectional alignment.
Because patch tokens encode localized time--frequency content, MiCLA applies hard projective residualization to eliminate components projected onto the counterpart representation.
In contrast, because the CLS token aggregates global clip-level context that may be shared across constituents, MiCLA employs sample-adaptive soft residualization to adjust the suppression ratio for each sample, avoiding the forced removal of shared acoustic and semantic context.
This dual-granularity design is intended to retain constituent-specific information while avoiding unnecessary removal of potentially shared global context from synthetic-mixture representations.

We evaluate MiCLA through supervised fine-tuning, frozen-encoder probing, and prototype retrieval.
In multi-label fine-tuning, MiCLA achieves 43.2\% mean average precision (mAP) on AudioSet-20K (AS-20K) and 50.3\% mAP on AudioSet-2M (AS-2M)~\cite{audioset-17}.
Under frozen Convex Gated Probing (CGP)~\cite{bat-26}, MiCLA attains 37.8\% mAP on AS-20K and 46.3\% mAP on AS-2M, achieving the highest performance among the evaluated models under the common CGP protocol on both benchmarks.
Additionally, in a Joint Semantic Retrieval diagnostic on synthetic ESC-50~\cite{esc50-15} mixtures, MiCLA improves simultaneous retrieval of both constituent classes compared with target-averaging baselines.
These results suggest that MiCLA provides an effective approach for polyphonic audio representation learning, while maintaining competitive general-purpose audio representation capabilities across diverse sound domains.

The primary contributions of this paper are summarized as follows:
\begin{itemize}
    \item We propose MiCLA, a self-supervised pre-training framework that leverages known mixture--constituent correspondence for bidirectional residual alignment.
    \item We instantiate this idea at patch and clip levels using hard projective residualization for patch tokens and sample-adaptive soft residualization for the global class token.
    \item We evaluate MiCLA across 11 downstream datasets and a Joint Semantic Retrieval diagnostic, demonstrating competitive transfer on polyphonic audio classification and improved joint constituent retrieval compared with target-averaging baselines.
\end{itemize}

\section{RELATED WORK}
This section reviews three areas related to MiCLA: masked audio pre-training, contrastive representation learning, and learning from audio mixtures.

\subsection{Masked Self-Distillation and Masked Audio Pre-training}
Masked audio pre-training divides a spectrogram into time--frequency patches and learns representations from partially observed inputs.
Reconstruction-based methods, including SSAST~\cite{ssast-22}, MAE-AST~\cite{mae-ast-22}, Audio-MAE~\cite{audiomae-22}, and MaskSpec~\cite{maskspec-23}, reconstruct masked spectrogram regions using vision transformer encoders such as the Audio Spectrogram Transformer (AST)~\cite{ast-21}.
In contrast, BEATs~\cite{beats-23} predicts discrete acoustic tokens.

Masked self-distillation instead uses a teacher network, often updated by an exponential moving average, to provide contextualized latent targets for a masked student.
This paradigm includes the modality-agnostic data2vec family~\cite{data2vec-22,data2vec2-23} and audio methods that combine masking, augmentation, and frame- or clip-level feature matching~\cite{eh-mam-24,atst-24,m2d-24,asit-24,eat-24,asda-25,matpac-25}.
In this paper, we use the term ``masked self-distillation'' for this teacher--student latent-prediction setting.
While existing masked self-distillation methods learn primarily from masked views of the same clip, MiCLA retains this same-clip training and introduces a cross-input signal relating a synthetic mixture to its two constituent clips.
For frozen evaluation, we adopt Convex Gated Probing (CGP), introduced in BAT~\cite{bat-26}, to compare encoders under a common probing recipe; BAT's front-end and attention modifications address a design axis distinct from MiCLA.

\subsection{Contrastive Structuring of Audio Representations}
Contrastive learning directly shapes similarities in the representation space.
Audio methods have combined contrastive objectives with momentum teachers, multi-scale features, clustering, or structured time--frequency views~\cite{mast-23,slicer-23,audiomosaic-26}.
AaSP~\cite{aasp-26}, which also builds on teacher--student masked modeling, uses aliasing-aware patch representations and multi-mask contrastive regularization to stabilize patch tokens across masked views.

MiCLA uses instance-level contrastive regularization (ICR) to structure visible patch representations before and during mixture alignment.
Its positives are different masked views of the same constituent clip, and its negatives are views of other constituent clips in the mini-batch; synthetic mixtures are excluded from the negative set.
Because unlabeled constituent clips may share sound events, acoustic scenes, or semantics, these negatives distinguish clip instances rather than semantic classes.
ICR thereby maintains the constituent-side geometry on which Stage~2 imposes mixture-to-constituent alignment.

\subsection{Learning from Mixtures and Polyphonic Audio}
Real-world recordings often contain multiple simultaneous sound events, and robustness to polyphonic inputs is important for audio classification~\cite{audioset-17}.
Mixup~\cite{mixup-18} on spectrograms or waveforms is commonly used during supervised fine-tuning to improve classifier robustness~\cite{m2d-24,eat-24,asda-25,audiomosaic-26,sslam-25}.
Such label-space regularization does not by itself constrain how a mixture representation relates to its constituent clips.
MiCLA instead retains that correspondence during self-supervised pre-training and uses mixing as a latent alignment signal rather than only as augmentation.

A closely related approach is SSLAM~\cite{sslam-25}, which also exploits the correspondence between mixed and pre-mixture spectrograms during self-supervised pre-training.
For masked patch prediction, SSLAM uses the average of the corresponding teacher targets from the two constituent inputs.
MiCLA replaces this averaged-target objective with directional residual alignment and also introduces constituent-specific alignment for CLS tokens.
It uses hard projective residualization for patch tokens and sample-adaptive soft residualization for CLS tokens.

Source separation methods, including permutation-invariant training~\cite{pit-17} and MixIT~\cite{mixit-20}, typically seek source-specific waveform estimates or time--frequency masks from physical mixtures.
MixIT, for example, trains a separator using mixtures of mixtures and constrains the estimated sources to explain the observed mixtures.
MiCLA reconstructs neither waveforms nor masks; its targets are representations of the full clips used in a controlled log-mel interpolation.
Accordingly, it addresses constituent-clip-aware representation learning, not signal separation.

MixMAE~\cite{mixmae-23} and MixedAE~\cite{mixedae-23} provide related examples in masked image modeling: they create mosaic-style mixed images by combining patches from different source images and use source correspondence for reconstruction or source recognition.
MiCLA applies the broader principle of source correspondence to audio but differs in the mixing mechanism.
Rather than replacing patches in a mosaic pattern, MiCLA performs time--frequency-bin-level convex interpolation over entire log-mel spectrograms and aligns teacher latent representations at both patch and clip levels.

\section{METHODOLOGY}
\label{sec:methodology}

This section presents Mixture-to-Constituent Latent Alignment (MiCLA).
First, Subsection~\ref{sec:preliminaries} introduces the self-supervised objectives used to establish foundational representations for isolated constituent clips.
Next, Subsection~\ref{sec:problem_formulation} formulates the challenges of polyphonic mixing and provides geometric motivation for directional residual alignment.
Finally, Subsection~\ref{sec:micla_framework} builds upon these geometric principles to describe the complete two-stage, dual-granularity MiCLA framework.

\subsection{Preliminaries}
\label{sec:preliminaries}

We first describe the self-supervised objectives used to learn representations of individual constituent clips.

\textbf{Masked Self-Distillation.}
An audio clip is converted into a log-mel spectrogram $\mathbf{x}\in\mathbb{R}^{F\times T}$.
Here, $F$ and $T$ denote the numbers of mel-frequency bins and time frames, respectively.
The spectrogram is divided into $N$ time--frequency patches, excluding the CLS token.
For each masked view, the patch indices are partitioned into a visible set $\mathcal{V}$ and a masked set $\mathcal{M}$.
These sets satisfy $\mathcal{V}\cap\mathcal{M}=\emptyset$ and $\mathcal{V}\cup\mathcal{M}=\{1,\ldots,N\}$.
For each clip, multiple masked views are generated by independently sampling a random mask pattern according to the inverse block masking scheme described in Section~\ref{sec:implementation_details}~\cite{data2vec2-23, eat-24}.
Let $f_{\theta}$ denote the student audio encoder~\cite{vit-21,ast-21}, and let $f_{\xi}$ denote its exponential moving average (EMA) teacher~\cite{data2vec-22}.
The student processes a masked view to output visible patch tokens $\mathbf{u}^{(j)}\in\mathbb{R}^{D}$ ($j\in\mathcal{V}$) along with a final-layer CLS representation $\mathbf{z}\in\mathbb{R}^{D}$.
Concurrently, the teacher processes the unmasked input to provide a patch target $\mathbf{p}^{(j)}\in\mathbb{R}^{D}$ at each patch position $j$ and a layer-normalized clip target $\mathbf{t}\in\mathbb{R}^{D}$.
To predict the masked regions, the visible student tokens are restored to their original positions, learnable mask tokens are inserted at positions in $\mathcal{M}$, and a predictor $g_{\mathrm{p}}$ estimates the corresponding teacher patch targets.
The masked self-distillation loss is defined as follows:

\begin{equation}
\ell_{\mathrm{base}}(\mathbf{x})
=
\frac{1}{|\mathcal{M}|}
\sum_{j\in\mathcal{M}}
\left\|
\hat{\mathbf{p}}^{(j)}
-
\operatorname{sg}
\left(
\mathbf{p}^{(j)}
\right)
\right\|_2^2
+
\left\|
\mathbf{z}
-
\operatorname{sg}
\left(
\mathbf{t}
\right)
\right\|_2^2 .
\label{eq:base_loss}
\end{equation}

Here, $\hat{\mathbf{p}}^{(j)}$ is the predictor output at position $j$.
The operator $\operatorname{sg}(\cdot)$ denotes stop-gradient.
The teacher targets are specified in Section~\ref{sec:implementation_details}.

\textbf{Instance-Level Contrastive Regularization.}
Following AaSP~\cite{aasp-26}, we apply instance-level contrastive regularization (ICR) to the visible patch representations.
The visible patch tokens $\{\mathbf{u}^{(j)}\}_{j\in\mathcal{V}}$ of each masked view are averaged and projected by an MLP $m_{\phi}$ to obtain an embedding $\mathbf{s}$.
Different masked views of the same clip form positive pairs, whereas views of other constituent clips in the mini-batch serve as negatives.
For an anchor representation $\mathbf{s}$, let $\mathbf{s}^{+}$ denote a positive representation and $\mathcal{N}^{-}$ its set of negative representations.
With temperature parameter $\tau$ and cosine similarity $\operatorname{sim}(\cdot,\cdot)$, we define $q^{+}=\exp(\operatorname{sim}(\mathbf{s},\mathbf{s}^{+})/\tau)$ and $q^{-}=\sum_{\mathbf{s}^{-}\in\mathcal{N}^{-}}\exp(\operatorname{sim}(\mathbf{s},\mathbf{s}^{-})/\tau)$.
The contrastive loss is then defined as follows:

\begin{equation}
\ell_{\mathrm{ctr}}
=
-\log
\frac{q^{+}}{q^{+}+q^{-}} .
\label{eq:contrastive_loss}
\end{equation}

The loss is averaged over the anchor--positive pairs in the mini-batch.
Together, $\ell_{\mathrm{base}}$ and $\ell_{\mathrm{ctr}}$ establish the constituent-side representation space used by MiCLA; however, extending these representations to polyphonic audio with overlapping sound events introduces critical geometric challenges, as formulated next.

\subsection{Problem Formulation and Motivation}
\label{sec:problem_formulation}

For multi-label classification of polyphonic audio, a mixture representation should preserve the overall acoustic context while retaining information associated with both constituents in a form accessible by downstream models.
We seek to learn this property without class labels by using the known correspondence between a synthetic mixture and its two constituent clips.

A naive approach is to align a single mixture representation $\mathbf{z}_{\mathrm{mix}}$ with both constituent targets $\mathbf{t}_a$ and $\mathbf{t}_b$.
With equal weighting, the objective $\|\mathbf{z}_{\mathrm{mix}}-\mathbf{t}_a\|_2^2+\|\mathbf{z}_{\mathrm{mix}}-\mathbf{t}_b\|_2^2$ is minimized at $\mathbf{z}_{\mathrm{mix}}=(\mathbf{t}_a+\mathbf{t}_b)/2$.
This optimum represents linear additivity up to scale.
However, it exists for any relative angle between the constituent targets and therefore provides no direct pressure to separate their directions.

Projective residualization adds such a separation pressure.
To isolate its geometric effect, consider an idealized setting in which each student constituent representation equals its teacher target, the predictor is the identity, and the targets are unit-normalized as $\|\mathbf{t}_a\|_2=\|\mathbf{t}_b\|_2=1$.
Let $\theta$ denote the angle between $\mathbf{t}_a$ and $\mathbf{t}_b$.
Removing the projection of $\mathbf{z}_{\mathrm{mix}}$ onto $\mathbf{t}_b$ yields $\mathbf{z}_{\mathrm{mix}}-(\mathbf{z}_{\mathrm{mix}}^{\top}\mathbf{t}_b)\mathbf{t}_b$, which is orthogonal to $\mathbf{t}_b$.
The residual error relative to $\mathbf{t}_a$ must therefore retain the component of $\mathbf{t}_a$ parallel to $\mathbf{t}_b$.
The minimum squared error is

\begin{equation}
\min_{\mathbf{z}_{\mathrm{mix}}}
\left\|
\mathbf{z}_{\mathrm{mix}}
-
(\mathbf{z}_{\mathrm{mix}}^{\top}\mathbf{t}_b)\mathbf{t}_b
-
\mathbf{t}_a
\right\|_2^2
=
(\mathbf{t}_a^{\top}\mathbf{t}_b)^2
=
\cos^2\theta .
\end{equation}

\noindent The reverse direction yields the same minimum error, with both directional minima achieved simultaneously at $\mathbf{z}_{\mathrm{mix}}=\mathbf{t}_a+\mathbf{t}_b$, giving a minimum bidirectional loss of $2\cos^2\theta$.
Under these idealized conditions, zero loss is achieved when target directions are orthogonal, while directional overlap incurs a strictly positive loss.
This analysis motivates residual alignment as a geometric pressure toward reduced directional overlap, rather than a guarantee of orthogonality in practical training.

MiCLA realizes this geometric pressure through bidirectional alignment.
In each direction, it suppresses the mixture component aligned with one constituent and aligns the residual with the teacher target of the other.
The counterpart representation remains on the differentiable student path, allowing the residual loss to shape the constituent geometry.
Reversing the target and counterpart roles applies this pressure to both constituent representations.

MiCLA applies the alignment at two granularities.
Patch tokens encode local time--frequency content and undergo full projection removal.
In contrast, because the CLS token summarizes global context that may be shared across clips, MiCLA employs bounded, sample-adaptive suppression ranging from no removal to full orthogonal residualization.
In practice, optimizing these alignment losses provides an effective inductive bias that promotes an additive and separated latent geometry during training, while soft residualization at the CLS level adaptively preserves shared context across constituent clips.

\subsection{The MiCLA Pre-Training Framework}
\label{sec:micla_framework}
Guided by the geometric motivation established above, we operationalize directional residual alignment into a concrete pre-training framework.
Below, we outline the two-stage training strategy, detail the dual-granularity alignment operations, and define the complete pre-training objectives.

\subsubsection{Overview: Two-Stage Pre-Training Strategy}
\label{sec:two_stage_strategy}

\begin{figure}[t]
\centering
\includegraphics[width=0.96\columnwidth]{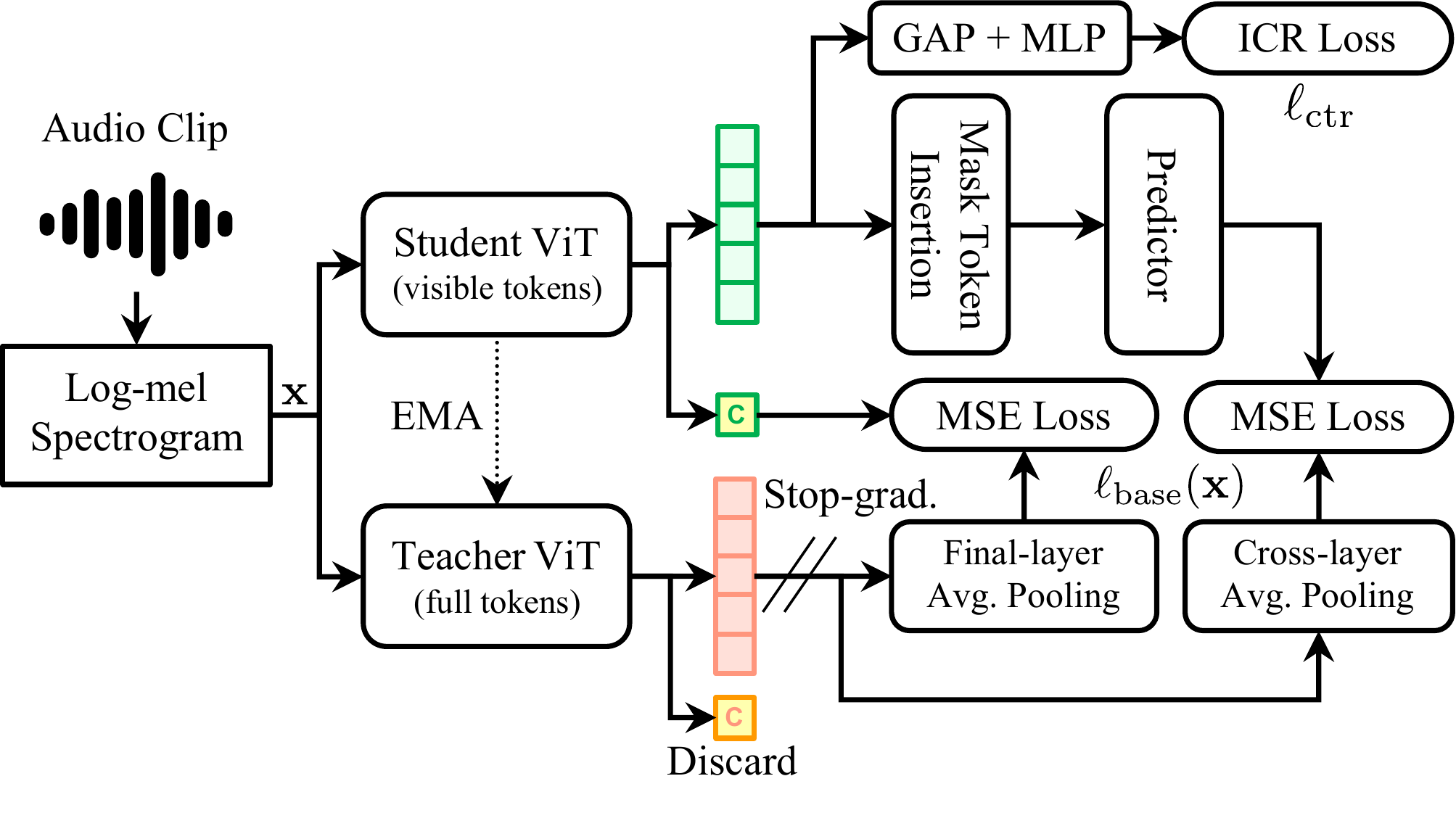}
\caption{%
Constituent-side pre-training in Stage~1.
Masked self-distillation optimizes $\ell_{\mathrm{base}}$.
Instance-level contrastive regularization $\ell_{\mathrm{ctr}}$ structures representations across masked views.
Both objectives are retained in Stage~2.
Boxes labeled ``C'' denote CLS tokens.
}
\label{fig:stage1_overview}
\end{figure}

MiCLA adopts a two-stage pre-training strategy following prior work~\cite{sslam-25}.
As illustrated in Fig.~\ref{fig:stage1_overview}, Stage~1 optimizes masked self-distillation and ICR using only constituent clips to establish a structured constituent-side representation space.

\begin{figure*}[t]
\centering
\includegraphics[width=0.98\textwidth]{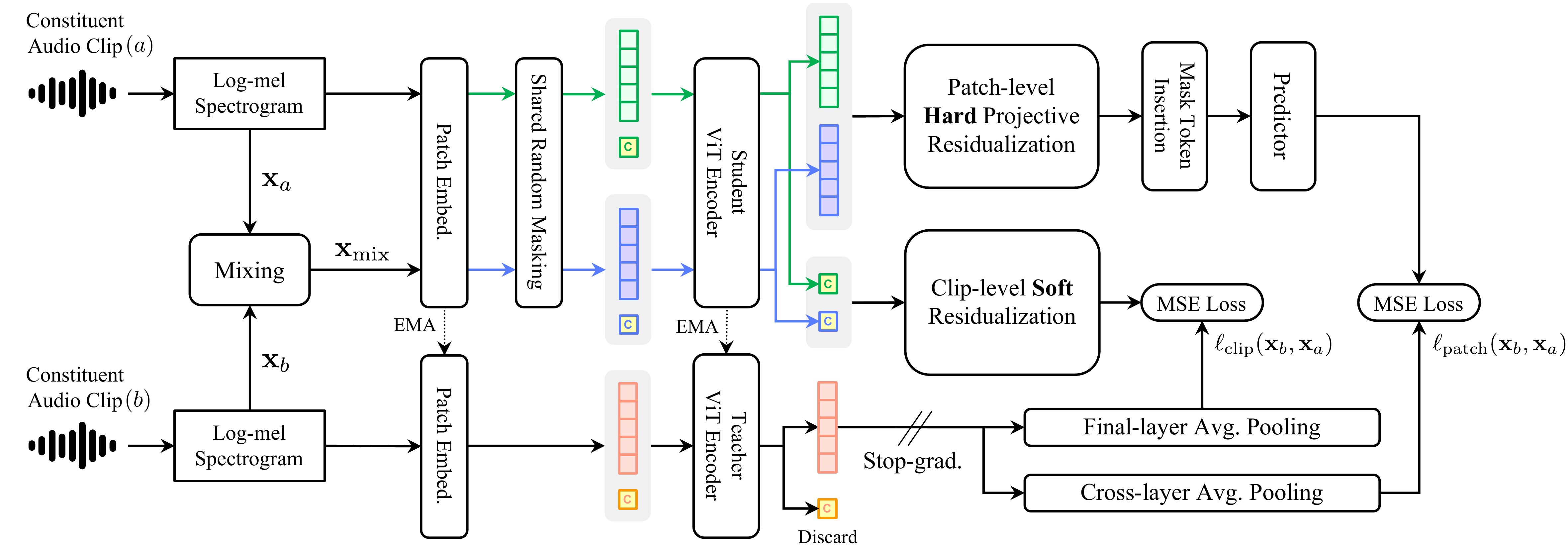}
\caption{%
Mixture-to-constituent latent alignment introduced in Stage~2.
In each target--counterpart direction, MiCLA suppresses the mixture component aligned with the counterpart.
The resulting residual is aligned with the target constituent's EMA-teacher representation.
Patch tokens use hard projective residualization.
CLS tokens use sample-adaptive soft residualization.
Boxes labeled ``C'' denote CLS tokens.
}
\label{fig:stage2_overview}
\end{figure*}

Stage~2 starts from the Stage~1 model and introduces mixture-to-constituent alignment, as shown in Fig.~\ref{fig:stage2_overview}.
Importantly, the constituent-side objectives are retained during Stage~2 to preserve the representation structure established in Stage~1 while introducing mixture-based alignment.

For each aligned view in Stage~2, the two constituent clips and their synthetic mixture share the same visible set $\mathcal{V}$ and masked set $\mathcal{M}$.
Shared random masking establishes token-wise correspondence at the same time--frequency positions.

\subsubsection{Dual-Granularity Mixture-to-Constituent Alignment}
\label{sec:dual_granularity_alignment}

Building upon the two-stage schedule and shared masking scheme, we now formalize the core directional alignment operations introduced in Stage~2.
Given the standardized log-mel spectrograms $\mathbf{x}_a$ and
$\mathbf{x}_b$ of two constituent clips, MiCLA constructs a synthetic
mixture by convexly interpolating them as follows:

\begin{equation}
\mathbf{x}_{\mathrm{mix}}
=
\mu\mathbf{x}_a
+
(1-\mu)\mathbf{x}_b .
\label{eq:mixture}
\end{equation}

\noindent The mixing coefficient is sampled as $\mu\sim\operatorname{Beta}(\eta,\eta)$.
The interpolation is applied after standardizing the constituent log-mel spectrograms.
It provides a controlled feature-space intervention with exact constituent correspondence.
It is not intended to model physical waveform mixing.

\textbf{Directional Alignment Setup.}
For the ordered pair $(\mathbf{x}_a,\mathbf{x}_b)$, $\mathbf{x}_a$ is the target and $\mathbf{x}_b$ is the counterpart to be suppressed.
The reverse direction $(\mathbf{x}_b,\mathbf{x}_a)$ exchanges these roles.
For latent vectors $\mathbf{u},\mathbf{v}\in\mathbb{R}^{D}$, we define the residualization operator as follows:

\begin{equation}
\rho_{\alpha}(\mathbf{u},\mathbf{v})
=
\mathbf{u}
-
\alpha
\frac{
\mathbf{u}^{\top}\mathbf{v}
}{
\|\mathbf{v}\|_2^2
}
\mathbf{v} .
\label{eq:residualization}
\end{equation}

\noindent Here, $\mathbf{u}$ is the representation to be residualized, $\mathbf{v}$ defines the counterpart direction, and $\alpha\in[0,1]$ controls the suppression strength.
Specifically, setting $\alpha=0$ leaves $\mathbf{u}$ unchanged, while setting $\alpha=1$ fully removes the orthogonal projection of $\mathbf{u}$ onto $\mathbf{v}$, yielding a hard residual orthogonal to $\mathbf{v}$ when $\mathbf{v}\neq\mathbf{0}$.
This residualization serves as a latent-space inductive bias without assuming that a learned mixture representation is the exact linear sum of its constituent representations.

\textbf{Patch-Level Hard Residualization.}
Because patch tokens capture localized time--frequency content, MiCLA fully removes the component aligned with the counterpart at each visible position.
Let $\mathbf{u}_{\mathrm{mix}}^{(j)}$ and $\mathbf{u}_{b}^{(j)}$ denote the student patch tokens of the mixture and counterpart at a shared visible position $j\in\mathcal{V}$.
For direction $(\mathbf{x}_a,\mathbf{x}_b)$, the hard residual is defined as $\mathbf{r}_{a|b}^{(j)}=\rho_{1}(\mathbf{u}_{\mathrm{mix}}^{(j)},\mathbf{u}_{b}^{(j)})$ for $j\in\mathcal{V}$.
These residual tokens are restored to their original positions, and learnable mask tokens are inserted at positions in $\mathcal{M}$.
The resulting sequence is processed by the same predictor $g_{\mathrm{p}}$ used for masked self-distillation, and the patch-level loss aligns each resulting output $\mathbf{h}_{a|b}^{(j)}$ with the target constituent's teacher representation $\mathbf{p}_a^{(j)}$ at masked position $j\in\mathcal{M}$.
The patch-level alignment loss is defined as follows:

\begin{equation}
\ell_{\mathrm{patch}}(\mathbf{x}_a,\mathbf{x}_b)
=
\frac{1}{|\mathcal{M}|}
\sum_{j\in\mathcal{M}}
\left\|
\mathbf{h}_{a|b}^{(j)}
-
\operatorname{sg}
\left(
\mathbf{p}_a^{(j)}
\right)
\right\|_2^2 .
\label{eq:patch_loss}
\end{equation}

\noindent Only the teacher target is detached.
Gradients propagate through both the mixture and counterpart student paths.

\textbf{Clip-Level Soft Residualization.}
The CLS token contains global context that may be shared by the two constituent clips, such as common acoustic environments or overlapping sound events.
Full orthogonal projection removal could discard this useful shared information.
MiCLA therefore selects the suppression strength adaptively for each sample.
Let $\mathbf{z}_{\mathrm{mix}}$, $\mathbf{z}_a$, and $\mathbf{z}_b$ denote the student CLS representations of the mixture, target constituent, and counterpart constituent, respectively.
For direction $(\mathbf{x}_a,\mathbf{x}_b)$, MiCLA selects the suppression strength from the student-side geometry by minimizing the squared distance between the residualized mixture representation and the target constituent's student representation $\mathbf{z}_a$:

\begin{equation}
\alpha_{a|b}^{\star}
=
\operatorname{sg}
\left(
\arg\min_{\alpha\in[0,1]}
\left\|
\rho_{\alpha}
\left(
\mathbf{z}_{\mathrm{mix}},
\mathbf{z}_b
\right)
-
\mathbf{z}_a
\right\|_2^2
\right) .
\label{eq:alpha_star_opt}
\end{equation}

\noindent As derived in Appendix~\ref{app:clip_derivation}, this student-side gate-selection problem has the following closed-form solution:

\begin{equation}
\alpha_{a|b}^{\star}
=
\operatorname{sg}
\left(
\operatorname{clip}_{[0,1]}
\left(
1
-
\frac{
\mathbf{z}_a^{\top}\mathbf{z}_b
}{
\mathbf{z}_{\mathrm{mix}}^{\top}\mathbf{z}_b
}
\right)
\right) .
\label{eq:alpha_star}
\end{equation}

\noindent Clipping restricts the operation to the range between no suppression ($\alpha=0$) and full projection removal ($\alpha=1$).
When $\mathbf{z}_{\mathrm{mix}}^{\top}\mathbf{z}_b$ is positive and non-negligible, a near-zero target--counterpart overlap $\mathbf{z}_a^{\top}\mathbf{z}_b$ yields $\alpha_{a|b}^{\star}\approx1$, whereas a larger positive overlap ratio reduces $\alpha_{a|b}^{\star}$, thereby retaining more counterpart-aligned latent content that may represent shared context.
Computing the gate entirely from student representations keeps all inner products in a common student latent coordinate system.
Applying the stop-gradient operator $\operatorname{sg}(\cdot)$ detaches $\alpha_{a|b}^{\star}$ from backpropagation; thus, $\mathbf{z}_a$ affects the forward gate value but receives no gradient through this gate-selection path, preventing target-side information from creating a direct optimization shortcut.
Within the subsequent residualization $\rho_{\alpha_{a|b}^{\star}}(\mathbf{z}_{\mathrm{mix}},\mathbf{z}_b)$, however, $\mathbf{z}_{\mathrm{mix}}$ and $\mathbf{z}_b$ remain differentiable.
Finally, let $\mathbf{t}_a$ denote the EMA-teacher clip target of the target constituent $\mathbf{x}_a$.
The residualized CLS representation is then aligned with the detached teacher target $\mathbf{t}_a$ via the clip-level alignment loss:

\begin{equation}
\ell_{\mathrm{clip}}(\mathbf{x}_a,\mathbf{x}_b)
=
\left\|
\rho_{\alpha_{a|b}^{\star}}
\left(
\mathbf{z}_{\mathrm{mix}},
\mathbf{z}_b
\right)
-
\operatorname{sg}
\left(
\mathbf{t}_a
\right)
\right\|_2^2 .
\label{eq:clip_loss}
\end{equation}

\noindent Unlike patch-level alignment, clip-level alignment operates directly on the representations without using the predictor.

\subsubsection{Pre-Training Objectives}
\label{sec:pretraining_objectives}

Having formulated the constituent-side and mixture-alignment losses, we now consolidate them into the complete optimization objectives for Stage~1 and Stage~2.

\textbf{Stage 1: Constituent-Side Objective.}
Stage~1 uses only constituent clips.
The expectation is taken over the Stage~1 training distribution, including sampled clips and masked views.
The Stage~1 objective is

\begin{equation}
\mathcal{L}_{\mathrm{stage1}}
=
\mathbb{E}
\left[
\lambda_{\mathrm{base}}
\ell_{\mathrm{base}}(\mathbf{x})
+
\lambda_{\mathrm{ctr}}
\ell_{\mathrm{ctr}}
\right] .
\label{eq:stage1_loss}
\end{equation}

\noindent Here, $\lambda_{\mathrm{base}}$ and $\lambda_{\mathrm{ctr}}$ weight masked self-distillation and ICR, respectively.

\textbf{Stage 2: Joint Mixture--Constituent Objective.}
Stage~2 retains the constituent-side objectives and adds bidirectional mixture-to-constituent alignment.
For a constituent pair $(\mathbf{x}_a,\mathbf{x}_b)$, the averaged base loss is $\bar{\ell}_{\mathrm{base}}=\frac{1}{2}\left[\ell_{\mathrm{base}}(\mathbf{x}_a)+\ell_{\mathrm{base}}(\mathbf{x}_b)\right]$.
For the fixed mixture, the directional loss is $\ell_{\mathrm{mix}}(\mathbf{x}_a,\mathbf{x}_b)=\ell_{\mathrm{patch}}(\mathbf{x}_a,\mathbf{x}_b)+\ell_{\mathrm{clip}}(\mathbf{x}_a,\mathbf{x}_b)$, where the first argument denotes the target and the second denotes the counterpart.
The reverse term uses the same mixture and mixing coefficient while exchanging only these roles.
The complete bidirectional loss is $\bar{\ell}_{\mathrm{mix}}=\frac{1}{2}\left[\ell_{\mathrm{mix}}(\mathbf{x}_a,\mathbf{x}_b)+\ell_{\mathrm{mix}}(\mathbf{x}_b,\mathbf{x}_a)\right]$.
The patch- and clip-level losses and the two directions are weighted equally.
Although the mixing ratio $\mu$ is used to interpolate the input spectrograms, it does not weight the alignment terms, encouraging both constituents to remain accessible even in imbalanced mixtures.
The expectation is taken over the Stage~2 training distribution, including constituent pairings, masked views, and mixing coefficients.
The Stage~2 objective is

\begin{equation}
\mathcal{L}_{\mathrm{stage2}}
=
\mathbb{E}
\left[
\lambda_{\mathrm{base}}
\bar{\ell}_{\mathrm{base}}
+
\lambda_{\mathrm{ctr}}
\ell_{\mathrm{ctr}}
+
\lambda_{\mathrm{mix}}
\bar{\ell}_{\mathrm{mix}}
\right] .
\label{eq:stage2_loss}
\end{equation}

\noindent Here, $\lambda_{\mathrm{mix}}$ weights the proposed alignment objective.
In Stage~2, $\ell_{\mathrm{ctr}}$ is computed only from constituent views, whereas synthetic mixture views contribute only to $\bar{\ell}_{\mathrm{mix}}$.

After pre-training, only the student encoder $f_{\theta}$ is retained for downstream tasks.
The EMA teacher $f_{\xi}$ is discarded.
The predictor $g_{\mathrm{p}}$ and projection head $m_{\phi}$ are also discarded.
Mixture construction and residualization are not used during downstream inference.
MiCLA therefore introduces no mixture-specific inference cost.

\section{EXPERIMENTS}
We evaluate MiCLA along four complementary dimensions:
(1) supervised fine-tuning to assess downstream adaptability on polyphonic multi-label classification;
(2) frozen-encoder probing via Convex Gated Probing (CGP) to measure multi-label transfer using layer- and token-wise features;
(3) fine-tuning and frozen CGP across single-label datasets to characterize transfer beyond the target multi-label setting; and
(4) Joint Semantic Retrieval to evaluate whether both known constituent classes remain jointly accessible from the frozen CLS representation of a synthetic mixture.

The classification benchmarks span two broad regimes: multi-label sound-event classification (AudioSet and FSD-50K), where clips may contain multiple overlapping acoustic events, and single-label classification across nine diverse speech, music, and environmental sound benchmarks.
In fine-tuning, the encoder and classification head are updated jointly to evaluate adaptability.
For frozen evaluation, the encoder remains fixed.
We measure representational accessibility primarily with CGP, which accesses features across layers and tokens, and use a final-layer linear probe as a supplementary diagnostic (Appendix~\ref{app:linear_probe}).

\subsection{Multi-Label Datasets}
\textbf{AudioSet}~\cite{audioset-17} is a YouTube-derived multi-label dataset with 527 sound-event classes.
\textbf{AS-2M} denotes the available unbalanced training split of approximately 1.96M clips, and \textbf{AS-20K} denotes the available balanced subset of approximately 21k clips; both use the same available evaluation set of approximately 19k clips and report mean average precision (mAP).

\textbf{FSD-50K}~\cite{fsd50k-21} is a Freesound-based multi-label sound-event dataset with 51,197 clips and 200 classes from the AudioSet ontology.
We use its official 36,796/4,170/10,231 train/validation/evaluation split for frozen probing and report mAP, thereby evaluating transfer beyond AudioSet's recording source and label distribution.

\subsection{Single-Label Datasets}
For single-label classification, we use the downstream datasets included in the EVAR\footnote{\url{https://github.com/nttcslab/eval-audio-repr}} platform.
Although EVAR is often used for linear probing, we use its dataset suite mainly for frozen CGP evaluation~\cite{bat-26}; fine-tuning performance on ESC-50 and SPC-2 is also reported in Table~\ref{tab:finetuning}.

\textbf{ESC-50}~\cite{esc50-15} is a single-label dataset for environmental sound classification.
It contains 2,000 five-second clips across 50 classes, with 40 clips per class; we follow the official five-fold cross-validation protocol and report average accuracy.

\textbf{UrbanSound8K}~\cite{us8k-14}, abbreviated as US8K, is a dataset for classifying urban sound events.
It contains 8,732 clips of up to four seconds across ten classes; we follow the official ten-fold split and report average accuracy.

\textbf{Speech Commands V2}~\cite{spcv2-18}, abbreviated as SPC-2, is a keyword spotting dataset for classifying short spoken commands.
It contains 105,829 approximately one-second utterances across 35 classes; the official split provides 84,843 training, 9,981 validation, and 11,005 test clips, and we report accuracy.

\textbf{VoxCeleb-1}~\cite{voxceleb1-17} is a large-scale speech dataset for speaker identification.
It contains 153,516 approximately eight-second utterances from 1,251 speakers; following prior work~\cite{m2d-24}, we use 138,361 training, 6,904 validation, and 8,251 test clips.

\textbf{VoxForge}~\cite{voxforge-18} is a speech dataset used for language identification.
It contains 176,428 approximately six-second utterances across six languages; following prior work~\cite{m2d-24}, we use 121,281 training, 26,684 validation, and 28,463 test clips.

\textbf{CREMA-D}~\cite{cremad-14} is a dataset for speech emotion recognition.
It contains 7,438 approximately 2.5-second utterances across six emotions; following prior work~\cite{m2d-24}, we use 5,155 training, 732 validation, and 1,551 test clips.

\textbf{GTZAN}~\cite{gtzan-02} is a dataset for music genre classification.
Following prior work~\cite{m2d-24}, we use 930 thirty-second music clips across ten genres, split into 443 training, 197 validation, and 290 test clips.

\textbf{NSynth}~\cite{nsynth-17} is a large-scale dataset used for instrument timbre classification.
It contains 305,979 four-second clips across 11 instrument families; the official split provides 289,205 training, 12,678 validation, and 4,096 test clips.

\textbf{Surge}~\cite{surge-21} is a pitch classification dataset generated by the Surge synthesizer.
It contains 183,392 four-second clips across 88 MIDI-note classes; following prior work~\cite{m2d-24}, we use 148,896 training, 17,160 validation, and 17,336 test clips.

\begin{table*}[t]
        \caption{Hyperparameters for Pre-training and Fine-tuning.}
\centering
{
\renewcommand{\arraystretch}{0.85}
\begin{threeparttable}
\begin{tabular}{l|c|c|cccc}
\toprule
\multirow{2}{*}{Hyperparameters} & Pre-Training (Stage 1) & Pre-Training (Stage 2) & \multicolumn{4}{c}{Fine-Tuning}  \\
 & AS-2M & AS-2M & AS-2M & AS-20K & ESC-50 & SPC-2 \\
\midrule
Optimizer & \multicolumn{6}{c}{AdamW~\cite{adamw-17} with $\beta_1 = 0.9, \beta_2 = 0.95$} \\
LR schedule & \multicolumn{6}{c}{Cosine Annealing~\cite{cossched-16} with linear warm-up} \\
Peak LR & $5.0\times10^{-4}$ & $5.0\times10^{-5}$ & \multicolumn{4}{c}{$5.0\times10^{-5}$} \\
Minimum LR & 0.0 & 0.0 & $1.0\times10^{-6}$ & $1.0\times10^{-6}$ & $1.0\times10^{-7}$ & $1.0\times10^{-7}$ \\
Layer decay~\cite{beit-21} & N/A & N/A & 0.65 & 0.80 & 0.75 & 0.75 \\
Weight decay & 0.05 & 0.05 & 0.05 & 0.05 & 0.05 & 0.05 \\
Training steps & 400K & 200K & 300K & 40K & 8K & 160K \\
Warm-up steps & 50K & 25K & 30K & 4K & 800 & 16K \\
Weighted sampling & N/A & N/A & 200K & N/A & N/A & N/A \\
Batch size per GPU & 64 & 64 & 32 & 32 & 128 & 128 \\
Masked views & 8 & 8,4\tnote{*} & N/A & N/A & N/A & N/A \\
\# of GPUs & 4 & 4 & 4 & 4 & 1 & 1 \\
Mixup~\cite{mixup-18} & N/A & N/A & 0.8 & 0.8 & 0.0 & 0.8 \\
Drop path~\cite{droppath-16} & 0.0 & 0.0 & 0.1 & 0.1 & 0.1 & 0.1 \\
\bottomrule
\end{tabular}
\begin{tablenotes}
        \footnotesize
        \item[*] In Stage 2, ``8,4'' denotes 8 constituent views per clip; 4 mixture views share masks with 4 of them. Mixture views enter only $\bar{\ell}_{\mathrm{mix}}$; all constituent views enter $\bar{\ell}_{\mathrm{base}}$ and $\ell_{\mathrm{ctr}}$, and matched ones also enter $\bar{\ell}_{\mathrm{mix}}$.
\end{tablenotes}
\end{threeparttable}
}
\label{tab:hyperparams_1}
\end{table*}

\subsection{Implementation Details}
\label{sec:implementation_details}
Unless otherwise stated, all audio data are resampled to 16 kHz.
Log-mel spectrograms are computed using a 400-sample Hann window, a hop length of 160 samples, 128 mel bins, and an FFT size of 1024.
Before encoder input, each log-mel spectrogram is standardized using a fixed mean of $-4.268$ and a standard deviation of $4.596$.
All experiments are conducted on NVIDIA H100 80-GB GPUs.
Due to the substantial computational cost of repeated pre-training, unless explicitly noted otherwise, downstream evaluations are conducted across three random seeds using a single shared pre-trained checkpoint, reporting the mean and standard deviation.

\subsubsection{Pre-Training Setup}
Table~\ref{tab:hyperparams_1} summarizes the hyperparameters used for both stages of pre-training.
Pre-training is performed on AS-2M.
AudioSet clips are approximately ten seconds long; following prior work~\cite{atst-24,m2d-24}, we randomly crop six-second segments for training efficiency.
We set the log-mel spectrogram length to 608 time frames by truncation or zero padding.

\textbf{Encoder and Positional Embeddings.}
The audio encoder is a post-norm 12-layer Vision Transformer~\cite{vit-21} with hidden dimension $D=768$.
Input spectrograms are partitioned into non-overlapping $16\times16$ time--frequency patches~\cite{eat-24}.
For positional representations, absolute positional embeddings are applied along the frequency axis to preserve absolute spectral positions, while Rotary Positional Embeddings (RoPE)~\cite{rope-24} are applied along the time axis to model relative temporal ordering.
In downstream tasks where input spectrogram length differs from pre-training, shorter inputs are processed directly, and longer inputs are extrapolated using YaRN~\cite{yarn-24}.

\textbf{Masking and Self-Distillation Architecture.}
Following EAT~\cite{eat-24}, we use $5\times5$ inverse block masking with a masking ratio of $80\%$.
Multi-mask training~\cite{data2vec2-23} applies multiple independently sampled mask patterns to the patch sequence of the same input spectrogram.
The EMA teacher parameters are updated with a momentum linearly scheduled from $0.9998$ to $0.99999$, warmed up across the entire duration in Stage~1 and for the first 100K steps in Stage~2~\cite{sslam-25}.
Patch-level teacher targets $\mathbf{p}^{(j)}$ are formed by averaging representations across all EMA teacher layers at patch position $j$.
The clip-level teacher target $\mathbf{t}$ is obtained by averaging all final-layer patch tokens from the EMA teacher and then applying layer normalization.
The patch predictor $g_{\mathrm{p}}$ is structured as a six-layer convolutional neural network (CNN) following EAT.

\textbf{Instance-Level Contrastive Regularization.}
The projection MLP $m_{\phi}$ consists of Linear, GELU, and Linear layers, with an intermediate hidden dimension of $D/2=384$.
The contrastive temperature is set to $\tau=0.2$.
The contrastive regularization weight is set to $\lambda_{\mathrm{ctr}}=0.01$ following AaSP~\cite{aasp-26}.

\textbf{Mixture Construction and Residualization.}
In Stage 2, synthetic mixtures are formed via round-robin pairing within each GPU mini-batch as $\mathbf{x}_{\mathrm{mix},i}=\mu_i\mathbf{x}_i+(1-\mu_i)\mathbf{x}_{i+1}$ with wrap-around indexing.
Each $\mu_i$ is sampled from $\operatorname{Beta}(\eta,\eta)$ with default parameter $\eta=4$.
For each pair, eight masked views are generated per constituent clip, four of which share mask patterns with four mixture views.
Mixture views enter only $\bar{\ell}_{\mathrm{mix}}$, whereas constituent views enter $\bar{\ell}_{\mathrm{base}}$ and $\ell_{\mathrm{ctr}}$ (with matched views also entering $\bar{\ell}_{\mathrm{mix}}$).
In all experiments, the squared-error losses $\ell_{\mathrm{base}}$ and $\ell_{\mathrm{mix}}$ are normalized by dividing by $\sqrt{D}$ before weighting.
We set loss weights $\lambda_{\mathrm{base}}=1.0$ and $\lambda_{\mathrm{mix}}=1.0$.
In projective residualization and clip gating, divisions by $\|\mathbf{v}\|_2^2$ and $\mathbf{z}_{\mathrm{mix}}^{\top}\mathbf{z}_{b}$ are stabilized with small sign-preserving offset ($10^{-6}$) to prevent numerical instability.

\subsubsection{Fine-Tuning Protocol}
Table~\ref{tab:hyperparams_1} summarizes the hyperparameters used for fine-tuning.
For fine-tuning evaluation, a classification head is attached to the CLS token of the pre-trained encoder.
The encoder and classification head are updated jointly using supervised data from each downstream task.
For augmentation, we employ Mixup~\cite{mixup-18} together with SpecAugment~\cite{specaug-19}.
In SpecAugment, the maximum time-mask width and frequency-mask width are each set to $20\%$ of the corresponding spectrogram dimension.
The ESC-50 experiments are conducted without any augmentation.
For all fine-tuning experiments other than ESC-50, we keep an exponential moving average of the model parameters, using a decay rate of $0.995$, to improve inference stability.
The training objective is chosen according to the dataset type.
Binary cross-entropy (BCE) with logits is used for multi-label classification.
Cross-entropy is used for single-label classification.
When Mixup is enabled for a single-label task, BCE is used instead of cross-entropy to handle soft targets.
Fine-tuning reflects not only the quality of the pre-trained representation but also its adaptability to each downstream task.

\begin{table*}[t]
        \caption{Hyperparameters for Convex Gated Probing (CGP).
The loss function is Asymmetric Binary Cross-Entropy (A-BCE)~\cite{asl-21} for multi-label datasets and Cross-Entropy (CE) for single-label datasets.}
\centering
{
\renewcommand{\arraystretch}{0.85}
\setlength{\tabcolsep}{4.25pt}
\begin{tabular}{l|cccccccccccc}
\toprule
\multirow{2}{*}{Hyperparameters}  & \multicolumn{12}{c}{Convex Gated Probing}  \\
 & AS-2M & AS-20K & FSD-50K & ESC-50 & US8K & SPC-2 & VoxCeleb-1 & VoxForge & CREMA-D & GTZAN & NSynth & Surge \\
\midrule
Optimizer & \multicolumn{12}{c}{AdamW~\cite{adamw-17} with $\beta_1 = 0.9, \beta_2 = 0.999$} \\
LR schedule & \multicolumn{12}{c}{Cosine Annealing~\cite{cossched-16} with linear warm-up}  \\
Peak LR & \multicolumn{12}{c}{$1.0\times10^{-3}$} \\
Minimum LR & \multicolumn{12}{c}{$1.0\times10^{-6}$} \\
Weight decay & 0.05 & 0.05 & 0.05 & 0.05 & 0.05 & 0.05 & 0.05 & 0.05 & 0.05 & 0.05 & 0.05 & 0.05 \\
Warm-up steps & 1K & 500 & 500 & 500 & 500 & 500 & 500 & 500 & 500 & 500 & 500 & 500 \\
Training steps & 200K & 25K & 25K & 25K & 25K & 25K & 25K & 25K & 25K & 25K & 25K & 25K \\
Weighted sampling & 200K & N/A & N/A & N/A & N/A & N/A & N/A & N/A & N/A & N/A & N/A & N/A \\
Batch size per GPU & 32 & 32 & 32 & 128 & 128 & 128 & 128 & 128 & 128 & 16 & 128 & 128 \\
\# of GPUs & 4 & 4 & 4 & 1 & 1 & 1 & 1 & 1 & 1 & 1 & 1 & 1 \\
Loss function & A-BCE & A-BCE & A-BCE & CE & CE & CE & CE & CE & CE & CE & CE & CE \\
\bottomrule
\end{tabular}
}
\label{tab:hyperparams_2}
\end{table*}

\subsubsection{Frozen Probing Protocol}
For our primary frozen-representation evaluation, we use Convex Gated Probing (CGP), proposed in BAT~\cite{bat-26}.
CGP is based on the observation that a standard final-layer linear probe, commonly used in image representation learning, may underestimate self-supervised learning (SSL) representations for audio.
In CGP, the encoder is frozen, and patch tokens and CLS tokens are extracted from each layer.
Layer-wise information is aggregated using learnable convex weights parameterized by a softmax function, and a classifier is trained using similarities to learnable prototypes.
This allows useful acoustic information contained not only in the final layer but also in intermediate layers to contribute to the evaluation.
We interpret CGP results as measuring the accessibility of information distributed across layers and tokens, rather than the linear separability of final-layer summary features.
The latter is separately examined by the supplementary linear probe in Appendix~\ref{app:linear_probe}.
Table~\ref{tab:hyperparams_2} summarizes the hyperparameters used for CGP.
Following BAT~\cite{bat-26}, the number of prototypes is set to 10,000.
For all CGP evaluations, no data augmentation is used.
We evaluate all baselines with the same CGP probing head and training recipe, strictly adhering to the authors' official data pre-processing pipelines in addition to using their official weights:
EAT~\cite{eat-24}\footnote{\url{https://huggingface.co/worstchan/EAT-base_epoch30_pretrain}},
SSLAM~\cite{sslam-25}\footnote{\url{https://huggingface.co/ta012/SSLAM_pretrain}},
AudioMosaic~\cite{audiomosaic-26}\footnote{\url{https://huggingface.co/hanxunh/AudioMosaic-vit-b16-pretrained}},
and BAT~\cite{bat-26}\footnote{\url{https://huggingface.co/lrauch/BAT-vit-b16-pretrainedAS2M}}.
For FSD-50K and GTZAN, we average chunk-level predictions, using each checkpoint's native length: $608$ frames for MiCLA and $1024$ frames for the other models.

As a supplementary diagnostic, we also conduct simple linear probing on the multi-label settings and report the results in Appendix~\ref{app:linear_probe}.
For this diagnostic, the encoder is frozen and only a linear classifier is trained.
The classifier input is constructed by concatenating the final-layer CLS token with the average of all final-layer patch tokens.
A sigmoid output is used for multi-label classification.
Unlike CGP, this probe does not use intermediate-layer aggregation or prototype-based readout.
We interpret it as a diagnostic of linear accessibility from final-layer summary features rather than as our primary frozen-probing protocol.

\subsection{Fine-Tuning Results}
Table~\ref{tab:finetuning} presents the fine-tuning results.
MiCLA demonstrates competitive transfer performance across multi-label benchmarks: among the models listed in Table~\ref{tab:finetuning}, MiCLA achieves 43.2\% mAP on AS-20K, improving upon prior baselines on this balanced benchmark.
On the full-scale AS-2M dataset, MiCLA achieves 50.3\% mAP, performing competitively with prior top-performing models (e.g., 50.2\% for SSLAM and AudioMosaic).

On single-label classification, MiCLA achieves 97.6\% accuracy on ESC-50, performing on par with or slightly above top baselines, while attaining competitive performance on SPC-2 ($98.1 \pm 0.1$\%).

\begin{table}[t]
\setlength{\tabcolsep}{2.8pt}
\caption{Fine-tuning results.
MiCLA results are reported as the mean $\pm$ standard deviation over three random seeds, while the results of the compared methods are cited as reported in the corresponding papers.}
\centering
\begin{tabular}{lccccc}
\toprule
\multirow{2}{*}{Model} & \multirow{2}{*}{\#Params.} & AS-2M & AS-20K & ESC-50 & SPC-2 \\
& & mAP (\%) & mAP (\%) & Acc (\%) & Acc (\%) \\
\midrule
MaskSpec~\cite{maskspec-23} & 86M & 47.1 & 32.3 & 89.6 & 97.7 \\
data2vec~\cite{data2vec-22} & 94M & - & 34.5 & - & - \\
Audio-MAE~\cite{audiomae-22} & 86M & 47.3 & 37.1 & 94.1 & 98.3 \\
M2D/0.7~\cite{m2d-24} & 86M & 47.9 & 38.6 & 96.0 & 98.4 \\
ASiT~\cite{asit-24} & 86M & 48.0 & 38.3 & 95.6 & 98.9 \\
BEATs$_{\mathrm{iter3}}$~\cite{beats-23} & 90M & 48.0 & 38.3 & 95.6 & 98.3 \\
ATST-Frame~\cite{atst-24} & 86M & 48.0 & 39.0 & - & 98.1 \\
EAT~\cite{eat-24} & 88M & 48.6 & 40.2 & 95.9 & 98.3 \\
ASDA~\cite{asda-25} & 93M & 49.0 & 41.5 & 96.1 & 98.3 \\
BAT~\cite{bat-26} & 93M & 48.9 & 41.6 & 96.0 & \textbf{99.1} \\
SSLAM~\cite{sslam-25} & 88M & 50.2 & 40.9 & 96.2 & 98.1 \\
AaSP~\cite{aasp-26} & 87M & 49.8 & 41.9 & 97.5 & 97.9 \\
AudioMosaic~\cite{audiomosaic-26} & 86M & 50.2 & 42.5 & 97.5 & 98.4 \\
\midrule
MiCLA~{\small (ours)} & 86M & \textbf{50.3}${\scriptstyle \pm 0.0}$ & \textbf{43.2}${\scriptstyle \pm 0.1}$ & \textbf{97.6}${\scriptstyle \pm 0.1}$ & 98.1${\scriptstyle \pm 0.1}$ \\
\bottomrule
\end{tabular}
\label{tab:finetuning}
\end{table}

\subsection{CGP Results on Multi-Label Datasets}
Table~\ref{tab:cgp_multilabel} compares frozen encoders under the common CGP protocol.
MiCLA achieves the highest performance among the compared models on both AudioSet benchmarks, obtaining $46.3 \pm 0.0$ mAP on AS-2M and $37.8 \pm 0.0$ mAP on AS-20K, exceeding the next-highest values by $0.6$ and $0.5$ points, respectively.
On the cross-dataset FSD-50K benchmark, MiCLA obtains $62.1 \pm 0.1$ mAP, improving upon several prior self-supervised models including EAT (59.5\%), SSLAM (57.8\%), and AudioMosaic (57.0\%), while performing close to BAT (62.5\%).

\begin{table}[t]
\setlength{\tabcolsep}{10pt}
\caption{CGP results on multi-label evaluation settings.
Values are reported as the mean $\pm$ standard deviation over three random seeds.}
\centering
\begin{tabular}{lccc}
\toprule
\multirow{2}{*}{Model} & AS-2M & AS-20K & FSD-50K \\
& mAP (\%) & mAP (\%) & mAP (\%) \\
\midrule
EAT~\cite{eat-24} & 45.0${\scriptstyle \pm 0.0}$ & 36.1${\scriptstyle \pm 0.0}$ & 59.5${\scriptstyle \pm 0.1}$ \\
SSLAM~\cite{sslam-25} & 43.8${\scriptstyle \pm 0.1}$ & 34.5${\scriptstyle \pm 0.0}$ & 57.8${\scriptstyle \pm 0.1}$ \\
AudioMosaic~\cite{audiomosaic-26} & 44.1${\scriptstyle \pm 0.0}$ & 35.0${\scriptstyle \pm 0.0}$ & 57.0${\scriptstyle \pm 0.3}$ \\
BAT~\cite{bat-26} & 45.7${\scriptstyle \pm 0.0}$ & 37.3${\scriptstyle \pm 0.1}$ & \textbf{62.5${\scriptstyle \pm 0.1}$} \\
\midrule
MiCLA~{\small (ours)} & \textbf{46.3${\scriptstyle \pm 0.0}$} & \textbf{37.8${\scriptstyle \pm 0.0}$} & 62.1${\scriptstyle \pm 0.1}$ \\
\bottomrule
\end{tabular}
\label{tab:cgp_multilabel}
\end{table}

\subsection{CGP Results on Single-Label Datasets}
Table~\ref{tab:cgp_singlelabel} shows the CGP results on single-label classification tasks.
MiCLA demonstrates strong transfer across diverse sound domains, ranking first on US8K ($92.5 \pm 0.1$\%) and NSynth ($86.4 \pm 0.1$\%), while placing second on SPC-2, GTZAN, and Surge.
While BAT incorporates specialized architectural and front-end modifications, MiCLA achieves competitive performance across several environmental and musical domains using a standard Vision Transformer backbone, highlighting the efficacy of its mixture-to-constituent pre-training objective.

\begin{table*}[t]
\setlength{\tabcolsep}{7.5pt}
\caption{CGP results on single-label classification tasks.
Values are reported as the mean $\pm$ standard deviation over three random seeds.}
\centering
\begin{tabular}{lccccccccc}
\toprule
\multirow{2}{*}{Model} & ESC-50 & US8K & SPC-2 & VoxCeleb-1 & VoxForge & CREMA-D & GTZAN & NSynth & Surge \\
& Acc (\%) & Acc (\%) & Acc (\%) & Acc (\%) & Acc (\%) & Acc (\%) & Acc (\%) & Acc (\%) & Acc (\%) \\
\midrule
EAT~\cite{eat-24} & 93.2${\scriptstyle \pm 0.1}$ & 89.9${\scriptstyle \pm 0.1}$ & 95.7${\scriptstyle \pm 0.0}$ & 66.3${\scriptstyle \pm 0.1}$ & 98.4${\scriptstyle \pm 0.2}$ & 74.6${\scriptstyle \pm 0.1}$ & 83.8${\scriptstyle \pm 0.2}$ & 83.5${\scriptstyle \pm 0.2}$ & 49.4${\scriptstyle \pm 0.2}$ \\
SSLAM~\cite{sslam-25} & 92.6${\scriptstyle \pm 0.1}$ & 88.7${\scriptstyle \pm 0.3}$ & 95.9${\scriptstyle \pm 0.1}$ & 64.0${\scriptstyle \pm 0.2}$ & 98.3${\scriptstyle \pm 0.2}$ & 74.9${\scriptstyle \pm 0.2}$ & 84.6${\scriptstyle \pm 0.2}$ & 82.3${\scriptstyle \pm 0.3}$ & 48.7${\scriptstyle \pm 0.0}$ \\
AudioMosaic~\cite{audiomosaic-26} & 91.0${\scriptstyle \pm 0.1}$ & 89.9${\scriptstyle \pm 0.1}$ & 95.9${\scriptstyle \pm 0.1}$ & 76.2${\scriptstyle \pm 0.2}$ & 97.7${\scriptstyle \pm 0.1}$ & 74.1${\scriptstyle \pm 0.3}$ & 83.4${\scriptstyle \pm 0.3}$ & 82.7${\scriptstyle \pm 0.1}$ & 47.9${\scriptstyle \pm 0.1}$ \\
BAT~\cite{bat-26} & \textbf{94.2${\scriptstyle \pm 0.1}$} & 91.2${\scriptstyle \pm 0.1}$ & \textbf{97.3${\scriptstyle \pm 0.0}$} & \textbf{81.7${\scriptstyle \pm 0.2}$} & \textbf{99.0${\scriptstyle \pm 0.1}$} & \textbf{76.0${\scriptstyle \pm 0.1}$} & \textbf{87.9${\scriptstyle \pm 0.0}$} & 84.2${\scriptstyle \pm 0.6}$ & \textbf{50.7${\scriptstyle \pm 0.1}$} \\
\midrule
MiCLA~{\small (ours)} & 92.5${\scriptstyle \pm 0.1}$ & \textbf{92.5${\scriptstyle \pm 0.1}$} & 96.6${\scriptstyle \pm 0.1}$ & 70.0${\scriptstyle \pm 0.1}$ & 98.1${\scriptstyle \pm 0.0}$ & 73.5${\scriptstyle \pm 0.2}$ & 85.6${\scriptstyle \pm 0.7}$ & \textbf{86.4${\scriptstyle \pm 0.1}$} & 49.6${\scriptstyle \pm 0.1}$ \\
\bottomrule
\end{tabular}
\label{tab:cgp_singlelabel}
\end{table*}

\begin{figure*}[t]
\centering
\includegraphics[width=\textwidth]{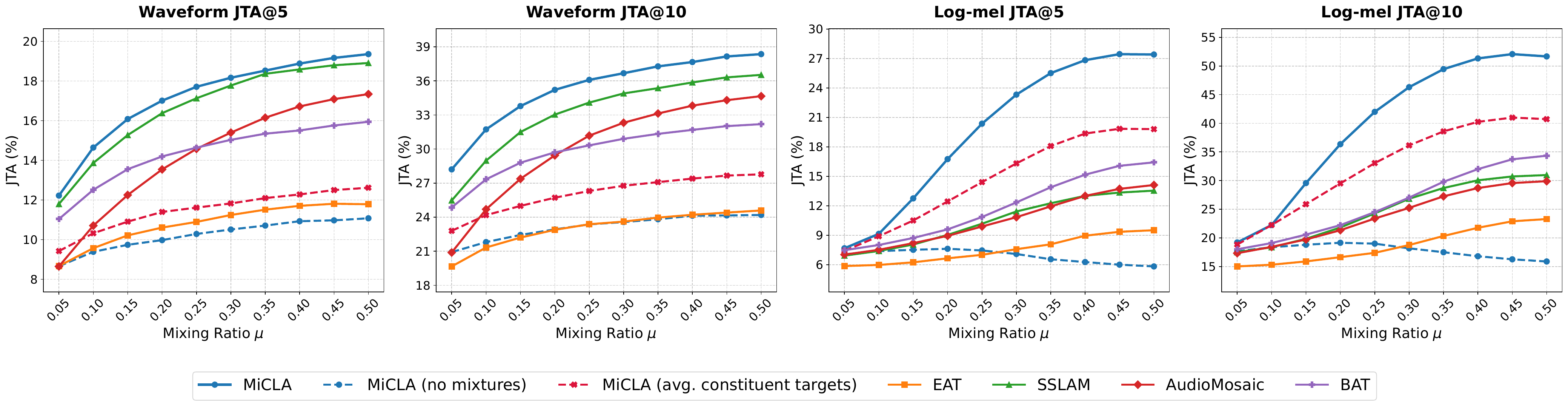}
\caption{Joint Semantic Retrieval on ESC-50.
The panels show JTA@5 and JTA@10 for waveform and log-mel mixtures across mixing ratios $\mu$.
A query is correct only when both constituent classes appear among the Top-$K$ retrieved classes.}
\label{fig:joint_retrieval}
\end{figure*}

\subsection{Joint Semantic Retrieval}
\label{sec:joint_semantic_retrieval}
Fine-tuning mAP does not directly show whether a mixture representation retains information about both constituents.
We conduct Joint Semantic Retrieval on ESC-50~\cite{esc50-15} to evaluate whether a frozen CLS representation of a synthetic mixture can retrieve both known constituent classes.

We compare MiCLA with two controlled ablations from Table~\ref{tab:ablation_stage2} using the same pre-trained checkpoints:
(1) ``MiCLA (no mixtures),'' which continues constituent-only training without $\ell_{\mathrm{mix}}$; and
(2) ``MiCLA (avg. constituent targets),'' a matched baseline that preserves the Stage-2 schedule, synthetic mixtures, and ICR, but replaces directional residual alignment with arithmetic averaging of constituent EMA targets at both patch and clip levels.
We also include external self-supervised baselines: EAT~\cite{eat-24}, SSLAM~\cite{sslam-25}, AudioMosaic~\cite{audiomosaic-26}, and BAT~\cite{bat-26}.
The ablations isolate the mixture-to-constituent signal, whereas external models provide broader context across different objectives and architectures.

For all models, the encoder is frozen.
For each fold and model, we average the $L_2$-normalized CLS embeddings of all unmixed training clips within each class and then $L_2$-normalize the resulting class prototype.

Evaluation uses clips from the held-out partition of each fold.
We consider all $\binom{50}{2}=1{,}225$ unordered class pairs.
For each class pair, we construct 10 mixture queries by pairing held-out clips from the two classes using a randomized ordering.
Clip instances may be reused across different class pairs and, when necessary, within a class pair.
This procedure yields $10\binom{50}{2}=12{,}250$ queries for each fold, model, mixing ratio $\mu$, and mixing domain.

For each query, the two clips are mixed with the specified mixing ratio.
Waveform mixtures are formed before spectrogram extraction.
Log-mel mixtures are formed in pre-normalization log-mel space before the remaining front-end processing.
We report $\mu$ from $0.05$ to $0.50$ because larger values are equivalent up to swapping the two constituent clips.

For each mixture, cosine similarities are computed between the mixture CLS token and the 50 class prototypes.
We evaluate Joint Top-$K$ Accuracy (JTA@$K$).
A retrieval is counted as correct only when both ground-truth constituent classes are included in the Top-$K$ retrieved classes.
For reference, random retrieval in this 50-class setting would achieve $\binom{5}{2}/\binom{50}{2}\approx0.82\%$ for JTA@5 and $\binom{10}{2}/\binom{50}{2}\approx3.67\%$ for JTA@10.

Figure~\ref{fig:joint_retrieval} shows JTA@5 and JTA@10 across mixing ratios $\mu$; each plotted point is averaged over the five folds.
MiCLA achieves the highest JTA across mixing ratios, mixture domains, and Top-$K$ settings.
All scalar comparisons below use the mean over all reported $\mu$ values after fold averaging.

For waveform mixtures, the averaged-target variant achieves $\mu$-averaged JTA@5/JTA@10 of $11.5\%/26.1\%$, modestly exceeding the no-mixture baseline ($10.2\%/23.1\%$).
Full MiCLA reaches $17.2\%/35.3\%$, improving over averaged targets by $5.7$ and $9.2$ percentage points, and also exceeding SSLAM ($16.7\%/33.2\%$).
For log-mel mixtures (the pre-training domain), the gains are larger: averaged targets achieve $14.7\%/32.6\%$ (vs. $6.9\%/17.8\%$ for no-mixture), while Full MiCLA reaches $19.7\%/40.0\%$, exceeding averaged targets by $5.0$ and $7.4$ points and performing higher than the evaluated baselines.

Across all four aggregate comparisons, performance consistently improves from the no-mixture baseline to averaged targets and Full MiCLA.
This progression indicates that while synthetic-mixture supervision alone aids joint retrieval, directional residual alignment provides clear additional gains.
Notably, whereas Full MiCLA and averaged targets differ by only $0.3$ mAP in AS-20K fine-tuning, their gap in Joint Semantic Retrieval reaches $5.0$--$9.2$ percentage points.
This contrast suggests that standard fine-tuning mAP may not fully expose the constituent-accessibility properties induced by pre-training objectives, which Joint Semantic Retrieval evaluates more directly.

\begin{table}[t]
\centering
\caption{Stage-1 ablation on constituent-clip representation structuring.
All rows use masked self-distillation and are evaluated by AS-20K fine-tuning mAP.
Values are the mean $\pm$ standard deviation over three random seeds.}
\label{tab:ablation_stage1}
\begin{threeparttable}
\footnotesize
\setlength{\tabcolsep}{10.0pt}
\begin{tabular}{lcccc}
\toprule
Ablation setting & ICR & RoPE & mAP & $\Delta$ \\
\midrule
No ICR              & \xmark & \cmark & 41.3${\scriptstyle \pm 0.1}$ & -1.3 \\
Absolute PosEmb.   & \cmark & \xmark & 41.4${\scriptstyle \pm 0.2}$ & -1.2 \\
Full Stage 1        & \cmark & \cmark & \textbf{42.6${\scriptstyle \pm 0.2}$} & 0.0 \\
\bottomrule
\end{tabular}
\begin{tablenotes}[flushleft]
\scriptsize
\item ``ICR'' denotes instance-level contrastive regularization over visible-token aggregates.
``RoPE'' denotes temporal rotary positional embedding.
``Absolute PosEmb.'' replaces temporal RoPE with absolute positional embedding.
``$\Delta$'' is relative to Full Stage 1.
\end{tablenotes}
\end{threeparttable}
\end{table}

\subsection{Ablation Study} \label{sec:ablation}
The ablations test how constituent-side structuring, mixture alignment, alignment granularity, soft clip-level residualization, the mixture construction domain, and the mixing-ratio distribution affect AS-20K fine-tuning mAP.
The component ablations report the mean and standard deviation over three random seeds, whereas the mixing-ratio sensitivity analysis in Table~\ref{tab:mixing_ratio} reports single representative runs.

\subsubsection{Stage-1 Constituent-Clip Structuring}
Table~\ref{tab:ablation_stage1} analyzes the Stage-1 components.
Both ICR and temporal RoPE improve the three-seed mean over their respective variants.
With masked self-distillation in every row, removing ICR lowers mAP from $42.6 \pm 0.2$ to $41.3 \pm 0.1$, while replacing temporal RoPE with absolute positional embedding yields $41.4 \pm 0.2$.

\begin{table}[t]
\centering
\caption{Stage-2 ablation on synthetic-mixture supervision and
mixture-to-constituent alignment.
All models are evaluated by AS-20K fine-tuning mAP.
Values are the mean $\pm$ standard deviation over three random seeds.}
\label{tab:ablation_stage2}
\begin{threeparttable}
\scriptsize
\setlength{\tabcolsep}{5.0pt}
\begin{tabular}{@{}lcccccccc@{}}
\toprule
Ablation setting & S2 & Mix & P & C & Soft & ICR & mAP & $\Delta$ \\
\midrule
\multicolumn{9}{l}{\textit{Comparison of Stage-2 training and mixture supervision}} \\
Stage-1 training
    & \xmark & \xmark & \xmark & \xmark & -- & \cmark
    & 42.6${\scriptstyle \pm 0.2}$ & -0.6 \\
Stage-2 no mixtures
    & \cmark & \xmark & \xmark & \xmark & -- & \cmark
    & 42.7${\scriptstyle \pm 0.1}$ & -0.5 \\
Avg. constituent targets
    & \cmark & \cmark & \cmark & \cmark & -- & \cmark
    & 42.9${\scriptstyle \pm 0.1}$ & -0.3 \\
\midrule
\multicolumn{9}{l}{\textit{Comparison of alignment granularities}} \\
Patch align. only
    & \cmark & \cmark & \cmark & \xmark & -- & \cmark
    & 42.9${\scriptstyle \pm 0.0}$ & -0.3 \\
Clip align. only
    & \cmark & \cmark & \xmark & \cmark & \cmark & \cmark
    & 43.1${\scriptstyle \pm 0.0}$ & -0.1 \\
\midrule
\multicolumn{9}{l}{\textit{Comparison of clip-level residualization variants}} \\
Hard clip removal
    & \cmark & \cmark & \cmark & \cmark & \xmark & \cmark
    & 43.0${\scriptstyle \pm 0.1}$ & -0.2 \\
\midrule
\multicolumn{9}{l}{\textit{Comparison with and without Stage-2 ICR}} \\
No ICR in Stage~2
    & \cmark & \cmark & \cmark & \cmark & \cmark & \xmark
    & 42.0${\scriptstyle \pm 0.1}$ & -1.2 \\
\midrule
Full MiCLA
    & \cmark & \cmark & \cmark & \cmark & \cmark & \cmark
    & \textbf{43.2${\scriptstyle \pm 0.1}$} & 0.0 \\
\bottomrule
\end{tabular}
\begin{tablenotes}[flushleft]
\scriptsize
\item ``S2'' denotes Stage~2 training.
``Mix'' denotes training with synthetic mixtures.
``P'' and ``C'' denote patch- and clip-level objectives.
``Avg. constituent targets'' uses the arithmetic mean of the two constituent
EMA targets at both levels, without residualization; the other mixture
variants use directional residual alignment.
``Soft'' denotes sample-adaptive clip residualization; \xmark~denotes
hard removal, and ``--'' indicates not applicable.
``ICR'' denotes instance-level contrastive regularization.
``$\Delta$'' is the difference in mean mAP relative to Full MiCLA.
\end{tablenotes}
\end{threeparttable}
\end{table}

\subsubsection{Stage-2 Mixture-to-Constituent Alignment}
Table~\ref{tab:ablation_stage2} organizes the Stage-2 ablations by design question.
Additional optimization alone produces only a minor gain: Stage-1 training achieves $42.6 \pm 0.2$ mAP, while continuing training without mixtures yields $42.7 \pm 0.1$.

To isolate directional residual alignment from mixture supervision itself, we evaluate a matched baseline (``Averaged constituent targets'').
This variant shares the Stage-2 schedule, synthetic mixtures, and ICR, but replaces directional residualization with arithmetic averaging of constituent EMA targets at both patch and clip levels.
Within this comparison, averaged targets achieve $42.9 \pm 0.1$ mAP ($+0.2$ over no mixtures), while Full MiCLA reaches $43.2 \pm 0.1$ ($+0.3$ over averaged targets).
In addition to this consistent fine-tuning gain, Section~\ref{sec:joint_semantic_retrieval} shows that directional residual alignment provides a clearer separation when directly evaluating joint constituent accessibility.

For alignment granularities, patch-only, clip-only, and joint alignment achieve $42.9 \pm 0.0$, $43.1 \pm 0.0$, and $43.2 \pm 0.1$ mAP, respectively, indicating that localized and global residual alignments provide complementary benefits to reach the highest performance in this comparison.
Similarly, sample-adaptive soft residualization achieves $43.2 \pm 0.1$ mAP compared with $43.0 \pm 0.1$ for hard removal, consistent with the intuition that adaptively retaining shared global context prevents over-suppression of common acoustic information.
Finally, removing Stage-2 ICR drops mAP to $42.0 \pm 0.1$, indicating that preserving constituent-side representation structure plays an important role during mixture alignment.

\subsubsection{Mixture Construction Domain}
Table~\ref{tab:mixing_domain} compares waveform, linear-mel, and log-mel domains for constructing synthetic mixtures while keeping the Stage-2 objective fixed.
Waveform mixing interpolates the two raw waveforms before front-end extraction, linear-mel mixing interpolates mel spectrograms before logarithmic compression, and log-mel mixing interpolates log-mel spectrograms as in Eq.~\eqref{eq:mixture}.
Log-mel mixing yields the highest AS-20K fine-tuning mAP.
This trend is consistent with the interpretation that log-mel interpolation provides a cleaner correspondence signal for alignment: it avoids the phase-dependent interference introduced by waveform mixing and, unlike linear-mel mixing, performs interpolation after dynamic-range compression, which may reduce the dominance of high-amplitude regions.

\begin{table}[t]
\centering
\caption{Ablation on the mixture construction domain.
Values are three-seed AS-20K fine-tuning mAP reported as mean $\pm$ standard deviation.}
\label{tab:mixing_domain}
\setlength{\tabcolsep}{6pt}
\begin{tabular}{lccc}
\toprule
Setting & Waveform & Linear-mel & Log-mel \\
\midrule
mAP (\%) & 42.9${\scriptstyle \pm 0.0}$ & 42.8${\scriptstyle \pm 0.1}$ & \textbf{43.2${\scriptstyle \pm 0.1}$} \\
\bottomrule
\end{tabular}
\end{table}

\subsubsection{Mixing-Ratio Sensitivity}
Performance varies little across the tested mixing-ratio distributions.
Table~\ref{tab:mixing_ratio} shows AS-20K fine-tuning mAP for $\eta\in\{1,2,4,6,8\}$, together with a deterministic mixing baseline using fixed $\mu=0.5$.
The highest single-run score occurs at $\eta=4$, while fixed $\mu=0.5$ also reaches $42.9\%$ mAP.
These results indicate that pre-training remains effective across various mixing-ratio distributions, supporting $\operatorname{Beta}(4,4)$ as a reliable default.

\begin{table}[t]
\centering
\caption{Sensitivity to the mixing-ratio distribution.
Values are AS-20K fine-tuning mAP from single representative runs.
``Fixed'' denotes deterministic mixing with $\mu=0.5$, whereas $\eta$ denotes the beta distribution parameter in $\mu\sim\operatorname{Beta}(\eta,\eta)$.}
\label{tab:mixing_ratio}
\setlength{\tabcolsep}{4pt}
\begin{tabular}{lcccccc}
\toprule
Setting & Fixed & $\eta=1$ & $\eta=2$ & $\eta=4$ & $\eta=6$ & $\eta=8$ \\
\midrule
mAP (\%) & 42.9 & 43.0 & 43.1 & \textbf{43.3} & 43.2 & 43.1 \\
\bottomrule
\end{tabular}
\end{table}

\section{CONCLUSION}
We presented MiCLA, a two-stage self-supervised pre-training method that uses synthetic mixture--constituent correspondence for polyphonic audio representation learning.
Stage~1 structures single-clip representations through masked self-distillation and contrastive regularization, while Stage~2 adds bidirectional alignment by residualizing a mixture against one constituent and aligning the residual with the other.
MiCLA applies hard projective residualization to local patch tokens and sample-adaptive soft residualization to the global class token.

Downstream evaluations show that MiCLA is particularly competitive in polyphonic multi-label settings, achieving $43.2\%$ mAP on AS-20K and $50.3\%$ on AS-2M under fine-tuning, as well as competitive scores on both AudioSet benchmarks under frozen Convex Gated Probing.
In Joint Semantic Retrieval, MiCLA improves mixing-ratio-averaged ($\mu$-averaged) JTA by $5.0$--$9.2$ percentage points over matched target-averaging supervision.
This retrieval gap, accompanied by consistent fine-tuning performance, indicates that direct retrieval diagnostics are particularly sensitive to the constituent-retention properties targeted by residual alignment.
Ablation studies support the design of the integrated MiCLA formulation, where dual-granularity alignment and constituent-structure preservation jointly contribute to downstream transfer.

MiCLA focuses on representation learning from pairwise log-mel mixtures rather than explicit source separation.
Future work includes evaluating on unconstrained real-world mixtures, unknown acoustic events, and mixtures with more than two sources.
Overall, our findings indicate that leveraging mixture--constituent correspondence is a promising direction for learning rich representations from polyphonic audio.

\appendices

\section{Derivation of Sample-Adaptive Clip-Level Soft Residualization}
\label{app:clip_derivation}

We derive the closed-form solution for the sample-adaptive suppression coefficient $\alpha_{a|b}^{\star}$ in Eqs.~\eqref{eq:alpha_star_opt} and~\eqref{eq:alpha_star}.
Let $\mathbf{z}_{\mathrm{mix}}$, $\mathbf{z}_a$, and $\mathbf{z}_b \in \mathbb{R}^D$ denote the student CLS representations of the mixture, target constituent $\mathbf{x}_a$, and counterpart constituent $\mathbf{x}_b$, respectively.

Consider a linear subtraction of the counterpart representation from the mixture, $\mathbf{z}_{\mathrm{mix}}-k\mathbf{z}_b$, parameterized by a scalar coefficient $k\in\mathbb{R}$.
To find the unconstrained coefficient $k_{a|b}^{\star}$ that minimizes the squared Euclidean distance to the target representation $\mathbf{z}_a$, we set up the objective:
\begin{equation}
J(k)
=
\left\|
\mathbf{z}_{\mathrm{mix}}
-
k\mathbf{z}_b
-
\mathbf{z}_a
\right\|_2^2 .
\end{equation}

\noindent Differentiating $J(k)$ with respect to $k$ yields
$\mathrm{d} J(k)/ \mathrm{d} k = -2 \mathbf{z}_b^{\top} (\mathbf{z}_{\mathrm{mix}} - k\mathbf{z}_b - \mathbf{z}_a)$.
Setting the derivative to zero, the unconstrained minimizer satisfies $(\mathbf{z}_{\mathrm{mix}} - k\mathbf{z}_b - \mathbf{z}_a) \perp \mathbf{z}_b$, yielding:

\begin{equation}
k_{a|b}^{\star}
=
\frac{
(\mathbf{z}_{\mathrm{mix}}-\mathbf{z}_a)^{\top}
\mathbf{z}_b
}{
\|\mathbf{z}_b\|_2^2
} .
\label{eq:app_k_star}
\end{equation}

\noindent Recall from Eq.~\eqref{eq:residualization} that the full orthogonal projection of $\mathbf{z}_{\mathrm{mix}}$ onto $\mathbf{z}_b$ corresponds to subtracting $k_{a|b}\mathbf{z}_b$ with

\begin{equation}
k_{a|b}
=
\frac{
\mathbf{z}_{\mathrm{mix}}^{\top}\mathbf{z}_b
}{
\|\mathbf{z}_b\|_2^2
} .
\label{eq:app_k_proj}
\end{equation}

\noindent When the residualization operator is expressed as $\rho_\alpha(\mathbf{z}_{\mathrm{mix}},\mathbf{z}_b)$ $= \mathbf{z}_{\mathrm{mix}} - \alpha k_{a|b}\mathbf{z}_b$, the unconstrained optimal suppression ratio becomes $\alpha = k_{a|b}^{\star}/k_{a|b}$.
Because both coefficients share the denominator $\|\mathbf{z}_b\|_2^2$, their ratio simplifies directly to:

\begin{equation}
\frac{k_{a|b}^{\star}}{k_{a|b}}
=
\frac{
(\mathbf{z}_{\mathrm{mix}}-\mathbf{z}_a)^{\top}\mathbf{z}_b
}{
\mathbf{z}_{\mathrm{mix}}^{\top}\mathbf{z}_b
}
=
1
-
\frac{
\mathbf{z}_a^{\top}\mathbf{z}_b
}{
\mathbf{z}_{\mathrm{mix}}^{\top}\mathbf{z}_b
} .
\end{equation}

\noindent Constraining $\alpha$ to the interval $[0,1]$ via clipping and applying stop-gradient $\operatorname{sg}(\cdot)$ yields the constrained solution in Eq.~\eqref{eq:alpha_star}:

\begin{equation}
\alpha_{a|b}^{\star}
=
\operatorname{sg}
\left(
\operatorname{clip}_{[0,1]}
\left(
1
-
\frac{
\mathbf{z}_a^{\top}\mathbf{z}_b
}{
\mathbf{z}_{\mathrm{mix}}^{\top}\mathbf{z}_b
}
\right)
\right) .
\end{equation}

\noindent As described in Section~\ref{sec:implementation_details}, the denominator $\mathbf{z}_{\mathrm{mix}}^\top\mathbf{z}_b$ is stabilized with a small sign-preserving offset ($10^{-6}$) to prevent numerical instability.

\section{Supplementary Linear Probing on Multi-Label Datasets}
\label{app:linear_probe}

\begin{table}[t]
\setlength{\tabcolsep}{10pt}
\caption{Supplementary linear probing results on multi-label settings.}
\centering
\begin{tabular}{lccc}
\toprule
\multirow{2}{*}{Model} & AS-2M & AS-20K & FSD-50K \\
& mAP (\%) & mAP (\%) & mAP (\%) \\
\midrule
EAT~\cite{eat-24} & 35.2 & 29.1 & 50.3 \\
SSLAM~\cite{sslam-25} & 33.2 & 27.4 & 47.2 \\
AudioMosaic~\cite{audiomosaic-26} & 36.2 & 30.9 & 47.6 \\
BAT~\cite{bat-26} & \textbf{37.4} & \textbf{32.5} & \textbf{55.8} \\
\midrule
MiCLA~{\small (ours)} & 34.7 & 29.5 & 52.1 \\
\bottomrule
\end{tabular}
\label{tab:linear_multilabel}
\end{table}

We additionally conduct simple linear probing on the multi-label settings as a supplementary diagnostic of linear accessibility.
This diagnostic is not used as the primary frozen-probing protocol in this paper.
For all models, the encoder is frozen and only a linear classifier is trained.
The classifier input is constructed by concatenating the final-layer CLS token with the average of all final-layer patch tokens.
Unlike CGP, this probe does not use intermediate-layer aggregation or prototype-based readout.

Table~\ref{tab:linear_multilabel} reports the supplementary linear probing results.
Under this simplified linear readout, MiCLA outperforms SSLAM across all three benchmarks ($+1.5$ to $+4.9$ mAP) and performs comparably to or higher than EAT on AS-20K and FSD-50K, while models with specialized front-ends or attention mechanisms, such as BAT, achieve higher performance under this protocol.
Comparing these results with the CGP evaluations in Table~\ref{tab:cgp_multilabel} suggests that a single final-layer linear classifier captures only part of the learned representations, and that multi-layer probing architectures can better access constituent-relevant information distributed across intermediate representations.

\bibliographystyle{IEEEtran}
\bibliography{references}

\end{document}